\RequirePackage{fix-cm}
\documentclass[smallextended]{svjour3}       % onecolumn (second format)
\smartqed  % flush right qed marks, e.g. at end of proof
\usepackage{graphicx}
\usepackage{amssymb}
\usepackage{graphics}
\usepackage{subfigure}

\usepackage{float}
\usepackage{enumitem}
\setlist[description]{leftmargin=\parindent,labelindent=\parindent}

\journalname{Preprint}

\begin{document}
\title{Towards experimental P-systems using multivesicular liposomes}
\titlerunning{Towards experimental P-systems}

\author{Richard Mayne \and Neil Phillips \and Andrew Adamatzky}

\institute{R. Mayne, N. Phillips \& A. Adamatzky \at
              Unconventional Computing Laboratory \\
              University of the West of England, Bristol, UK\\
              Tel.: +44 117 32 87861\\
              \email{richard.mayne@uwe.ac.uk}                   
}

\date{Received: date / Accepted: date}

\maketitle

\begin{abstract}
P-systems are abstract computational models inspired by the phospholipid bilayer membranes generated by biological cells. Illustrated here is a mechanism by which recursive liposome structures (multivesicular liposomes) may be experimentally produced through electroformation of dipalmitoylphosphatidylcholine (DOPC) films for use in `real' P-systems. We first present the electroformation protocol and microscopic characterisation of incident liposomes towards estimating the size of computing elements, level of internal compartment recursion, fault tolerance and stability. Following, we demonstrate multiple routes towards embedding symbols, namely modification of swelling solutions, passive diffusion and microinjection. Finally, we discuss how computing devices based on P-systems can be produced and their current limitations.
\end{abstract}


\section{Introduction}
\label{sec1}

A P-system or `membrane computer' is an abstract computational model inspired by biological reactions occurring within the confines of phospholipid (PL) membrane-encapsulated living cells and their subcomponents. Briefly, P-systems are envisaged as hypothetical recursive membrane-bound compartments wherein internal symbolic objects, likened to chemicals and catalysts, may react with each other to do computation, the products of which may remain within their compartment or diffuse inwards or outwards through their container's membrane. Introduced by P\u{a}un in 1998 \cite{DBLP:journals/eatcs/Paun99a,DBLP:journals/ijfcs/Paun00,Paun2002} (see comprehensive overview and analysis in \cite{Paun09,Paun2002introduction,Paun2010handbook}), several classes of P-systems have been proposed and applied to a range of fields including modelling of biological processes, cryptography, optimisation and control of robot swarms~\cite{Paun2006,paun2005further,frisco2014applications,florea2017membrane,zhang2017real}

P-systems have been considered as exclusively theoretical tools. In 2008 \cite{gershoni2008research} a pathway to a potential laboratory implementation was outlined which was aiming to compute the Fibonacci sequence in laboratory prototypes of P-systems. The compartmentalisation was proposed to be realised by test tubes, multisets implemented by DNA molecules, bio-inspired evolution rules executed by enzyme drive operations with the DNA molecules. Other key functionalities to be targeted were synchronisation of compartment evolution (this was proposed via delays) and output interfaces realised via a spectrophotometry. The theoretical discussions presented in  \cite{gershoni2008research}  led to a patent on a theoretical implementation of P-systems in a system of cascading test tubes~\cite{keinan2009membrane}. 

In this paper, we present laboratory experimental data demonstrating the basis by which architecture of `real' P-systems may be created. More specifically, we demonstrate the creation of recursive, synthetic microscopic membrane systems --- liposomes --- and detail progress towards describing computation within.

\begin{figure}
    \centering
    \subfigure[]{\includegraphics[width=0.55\textwidth]{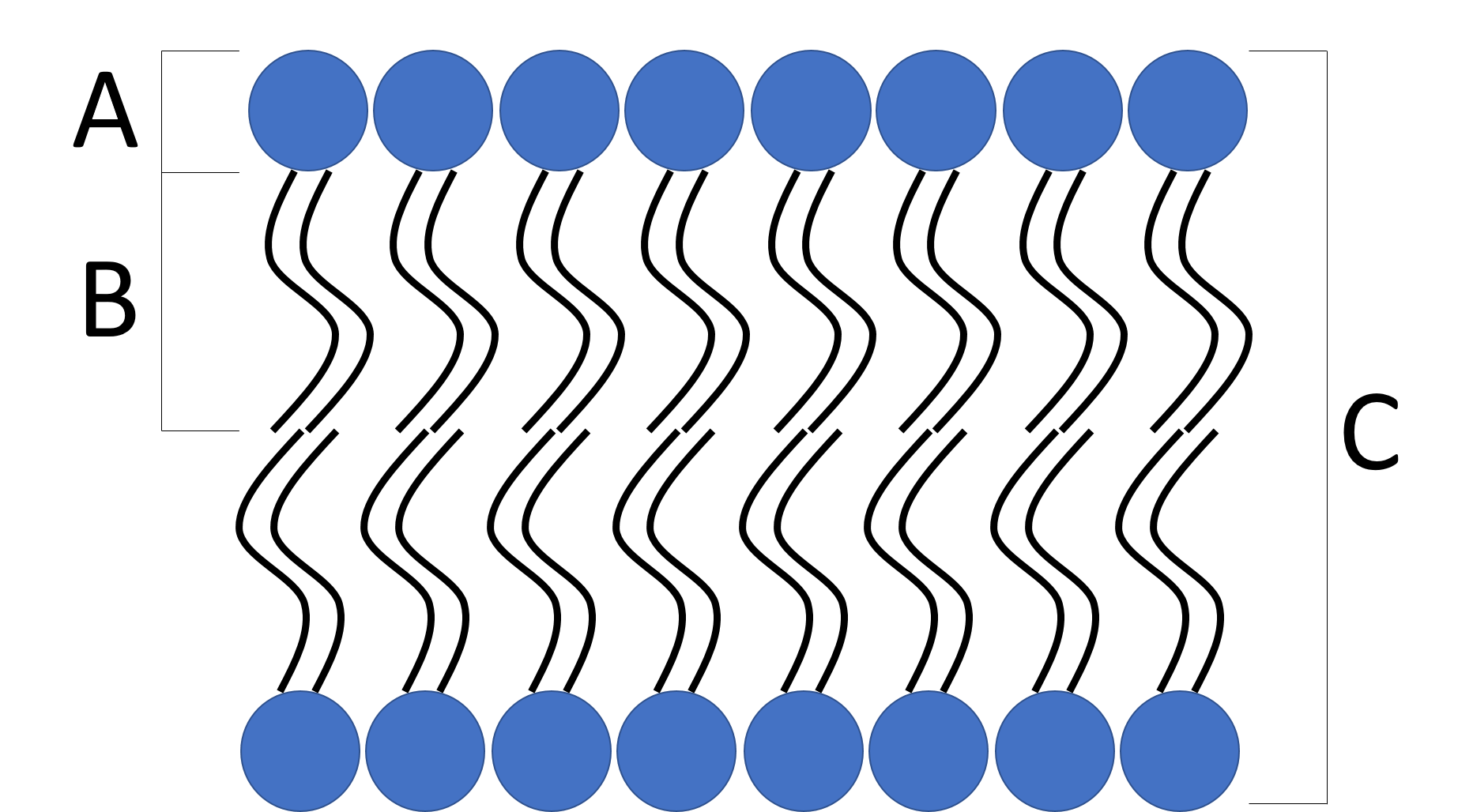}}
    \subfigure[]{\includegraphics[width=0.30\textwidth]{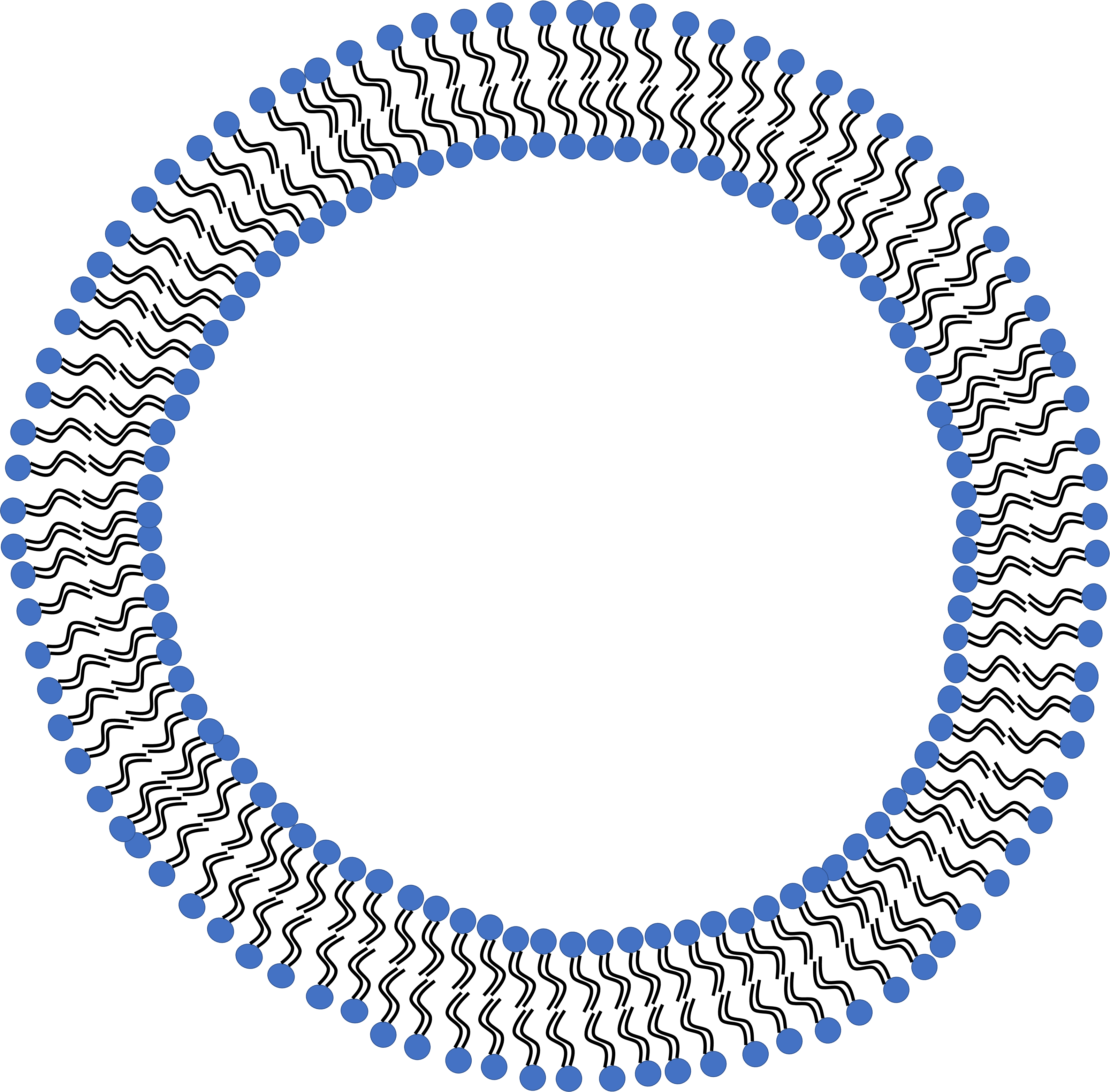}}
    \caption{Diagrams to demonstrate structure of a phospholipid bilayer with no accessory proteins. (a) Flat section of bilayer. A: hydrophilic head of the molecule; B: hydrophobic tail; C: bilayer unit. (b) Arrangement of phospholipds in a liposome. }
    \label{fig-bilayer}
\end{figure}

A liposome is defined a spherical compartment containing fluid that is comprised of at least one phospholipid (PL) bilayer, i.e. two layers of PL molecules, each incorporating a hydrophilic head and hydrophobic tail, arranged tail-to-tail (Fig. \ref{fig-bilayer}). Animal cell membranes are also formed from PL bilayers, meaning that liposomes and animal cells have similar properties; `natural' liposomes are synthesised by animal cells where they are used for substance storage and transport, but `synthetic' liposomes may also be produced in a laboratory from a variety of PLs and methods. Research into synthetic vesicle generation is an active area of investigation due to its relevance to medicine. 

The use of liposomes for computing applications is not novel: theoretical work on their use as minimal biological structures has been explored by a number of groups, both as modelling P-systems \cite{Smaldon2010} and `protocells' for the implementation of synthetic biological circuits \cite{Rasmussen2011,Stano2015,Kurihara2015}, although experimental studies in these fields are sparse.

Our previous excursion into experimental laboratory implementation of P-systems, without using the term, is two-fold. First, in 2011-2012 we implemented an analog of P-systems with Belousov-Zhabotinsky (BZ) vesicles, which built on earlier theoretical work on oscillating chemical reactions as examples of multiset computing systems by Suzuki {\it et al.} \cite{Suzuki2001} and then Manca {\it et al.} \cite{Manca2005}. Therein, a gas free analogue of the BZ reaction catalysed by ferroin was encapsulated in phospholipid stabilised vesicles: a reaction mixture which  exhibits spontaneous oscillation and excitation transfer between vesicles was demonstrated~\cite{costello2012initiation}. The prototypes realised compartmentalisation (BZ mixture inside liposomes), signal propagation between several BZ vesicles and optical output. We have demonstrated how ensembles BZ vesicles can implement logical circuits~\cite{adamatzky2012architectures,holley2011computational,adamatzky2011towards,holley2011logical} and some tasks of computational geometry~\cite{adamatzky2011vesicle}. The BZ-vesicles prototypes of P-system did not demonstrate embedding of one vesicle into another. Second, in 2015 we proposed that liposomes within live cells are a viable medium for implementing unconventional, `collision-based' logic as a route towards cellular reprogramming \cite{Mayne2015}. A cell can be seen as a P-system of first order (outer compartment) and liposomes as P-systems of second order (inner compartment). 

Here we describe the generation of a specific variety of liposome, the `multivesicular liposome' (MVL), which possess a single external PL bilayer and multiple internal liposomes. MVLs were first synthesised in 1983 \cite{Kim1983} and may be used as time-release drug delivery systems \cite{Kim1993,Dai2006}. Although exact formation protocols vary greatly between authors, MVLs are typically produced by a two-step emulsification process and may be engineered to contain a range of compounds such as proteins and nucleic acids. MVLs produced by these methods typically have a mean size in the region of 20--40~$\mu$m in diameter but may range to significantly larger or smaller than this and the number of internal compartments is also highly variable \cite{Kim1983,Yao2012}. It has been suggested that MVLs are intermediary products in the production of unilamellar, single compartment liposomes \cite{Yao2012}.

We present here different methods by which MVLs may be produced, namely electroformation on indium tin oxide (ITO) coated glass. Electroformation is a common method for producing standard, single compartment giant liposomes~\cite{angelova1986liposome,angelova1988mechanism,kuribayashi2006electroformation,estes2005giant,angelova2000liposome,patil2014novel,meleard2009giant,dao2017membrane,pereno2017electroformation}, also known as `giant unilamellar vesicles', but has not, to our knowledge, been described as a viable method for producing MVLs. The benefits of this approach are twofold:
\begin{enumerate}
    \item Electroformation is significantly simpler and cheaper than double emulsion methods, hence this represents a far more accessible technology for interdisciplinary Computer Science research.
    \item MVLs may be observed whilst they are forming, hence a certain degree of control may be exerted over their characteristics by altering various paramaters as they form.
\end{enumerate}

We proceed to characterise the MVLs produced through these methods, after which we present our findings on the immediate possibilities of implementing experimental P-systems, along with the remaining challenges.

\section{Methods}

MVLs were generated through electroformation on ITO-coated glass microscope slides, each with a resistance of 59~$\Omega$ and measuring $25 \times 75 \times 1.1$~mm, using dipalmitoylphosphatidylcholine (DOPC, Avanti Polar Lipids, USA) as the lipid substrate. Lipid solutions were prepared in chloroform at a concentration of 1~mg/ml and stored at -20$^o$C under a layer of nitrogen.

\begin{figure}
    \centering
    \subfigure[]{\includegraphics[width=0.40\textwidth]{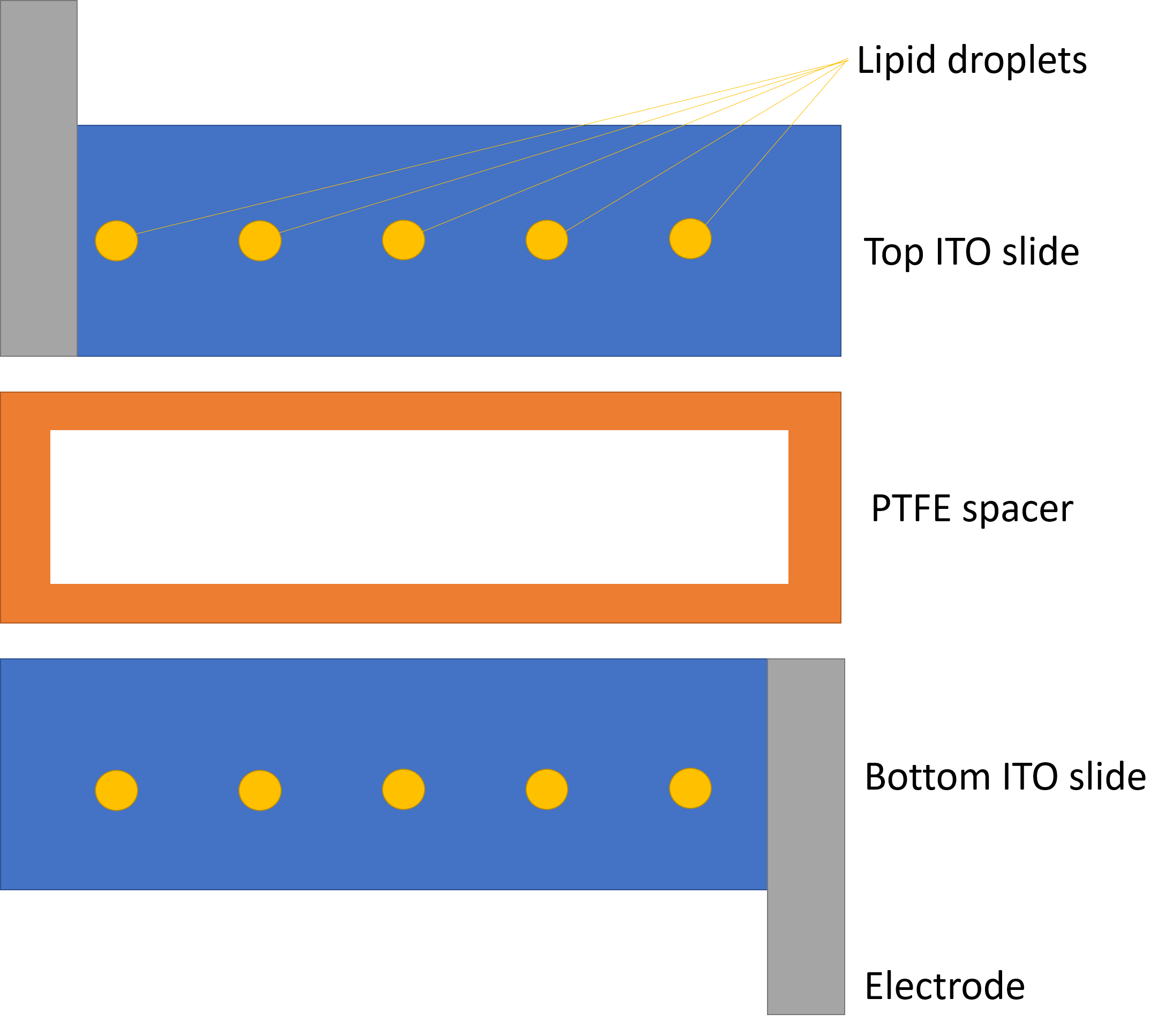}}
    \subfigure[]{\includegraphics[width=0.59\textwidth]{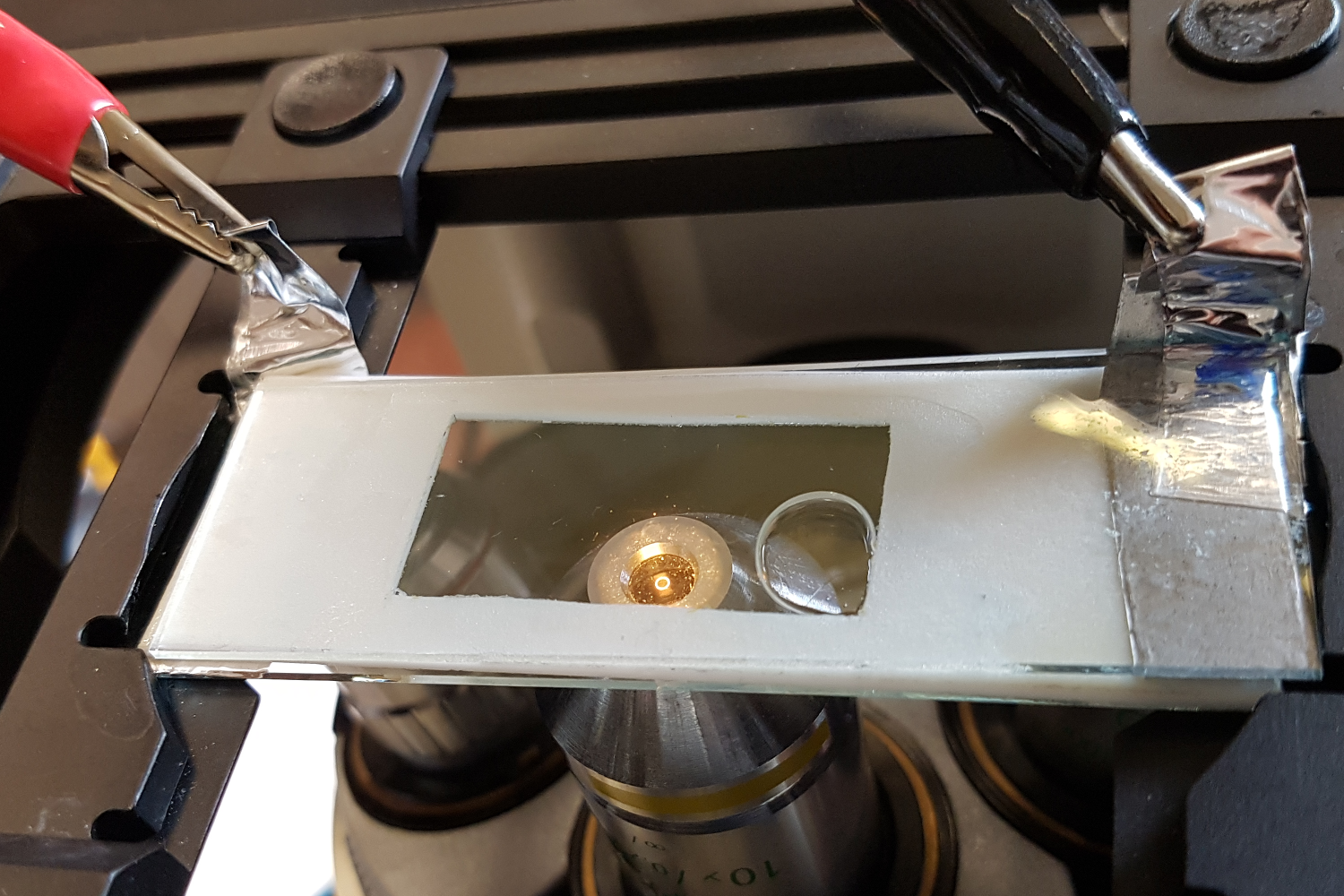}}
    \subfigure[]{\includegraphics[width=0.7\textwidth]{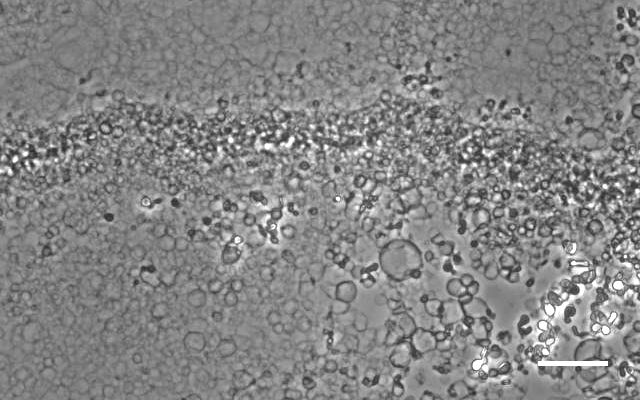}}
    \caption{Images to show construction of electroformation chamber. (a) Schematic to show the 3 layers of each chamber (not to scale). (b) Constructed chamber {\it in situ} on microscope, undergoing stimulation. (c) Example field of view after 40 minutes of agitation under 10~Hz. The vast majority of liposomes have no recursive contents. Scale bar 50~$\mu$m.}
    \label{fig-chamber}
\end{figure}

The electroformation chamber (Fig.~\ref{fig-chamber}) was constructed as follows: two ITO-coated slides were cleaned using ethanol and Whatman, grade 105 lens cleaning tissue (GE Healthcare, UK). Both slides then had a $90 \times 10 \times 0.07$~mm adhesive aluminium tape electrodes stuck to their edge perpendicular to the longer aspect of the ITO-coated slides. 

A polytetrafluoroethylene (PTFE) spacer of the same dimensions as the slides, which had a $20 \times 65$~mm void oriented centrally, was coated with double-sided adhesive tape on both sides and stuck to one of the slides. Both slides were transferred to a large sealed Petri dish which had a constant stream of nitrogen gas flowing across it at a rate of approximately 0.12~L/s. 20~$\mu$l of lipid solution was dropped onto each slide using a capillary tube in 5 individual 4~$\mu$l droplets. The lipid solution was left to evaporate under the stream of nitrogen for 2 hours. %Prev figure was 1.8l/s...where did that come from??

After the evaporation phase, the void delineated by the PTFE spacer was filled with approximately 1.3~mL solution of 300~mM sucrose in ultra-pure ($> 18.2$ M$\Omega$/cm) water (the `swelling solution'). Care was taken whilst dispensing the sucrose solution to not delaminate the dried lipid deposits. The other slide was then stuck to the spacer, thereby sealing the chamber. 

The electroformation chamber was transferred to a Zeiss Axiovert 200M inverted light microscope so that observations could be made whilst liposomes were being formed. The chamber's electrodes were connected to a BK4053B function generator (BK Precision, USA) set to 50~$\Omega$ output load. The chamber was initially stimulated with a 1.2~Vpp, 10~$\mu$A RMS sine wave at 10~Hz to swell the liposomes from the slides, after which the frequency was reduced to 2~Hz for a further hour to encourage their budding off. The timings for each phase were determined through regular microscopic observations of the liposomes as they formed. The same function generator was later used to stimulate liposome formations with DC pulses. %% More?

Observations were performed throughout the process at $\times$100--400 magnification using phase contrast optics. Photo and videomicrography was performed using a 340M-USB camera system (Thorlabs, USA). Video footage was captured at 60--400 frames per second, depending on the application. Measurements were made manually as ascertaining whether MVLs contained smaller vesicles required manual focus adjustment. Measurements were calculated using GNU Image Manipulation Program 2.8.22 measurement tool. The number of recursive liposomes were counted manually from still images. Manual counting was chosen because attempts to automate the process (via Circular Hough Transform) were found to be significantly more error prone. This was because the imaging technique (phase contrast), which renders edges in a gradient between black and white depending on their position relative to the field of view, was found to miss circles whose outlines were a similar colour to the background.

For liposome longevity studies, samples were gently pipetted from the electroformation chambers in their sucrose medium and stored in sealed glass bijou jars at 4$^o$C. Samples were checked every day by pipetting 5$\times$50~$\mu$l of liposome solution from each sample onto a glass well slide and manually checking via light microscopy; stability was assessed as having reduced when a MVL was not found in any of the droplets examined.

Micro-scale objects --- starch-coated magnetite (iron II/III oxide) nanoparticles (chemicell GmbH, Germany) with an original 200~nm diameter that tended to form clumps in the order of several microns --- were embedded into the liposomes generated through adding them to the swelling solution at a concentration of 1/100 (w/v). 

%{\bf Microsuction and microinjection of chemicals and objects into liposomes was performed using a CellTram Oil system (Eppendorf, Germany).}

\section{Results}
\subsection{MVL properties}
MVLs were observed to form on all areas of the slide that had been treated with DOPC solution, but were more likely to form in the slight ridges observed in dried lipid spots where the concentration of adherent lipid layers was presumably greater. Standard non-recursive liposomes were also produced {\it en masse} and were much more prevalent than MVLs, although the majority of them remained adherent to the surface of the ITO-coated slides. MVLs tended to be more prevalent 30--60 minutes of electroformation at 10~Hz but continued to emerge during the first 120 minutes of stimulation.

\begin{figure}[h!]
    \centering
   \subfigure[]{\includegraphics[width=0.49\textwidth]{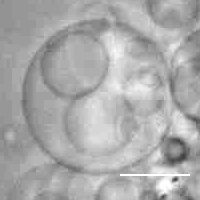}}
   \subfigure[]{\includegraphics[width=0.49\textwidth]{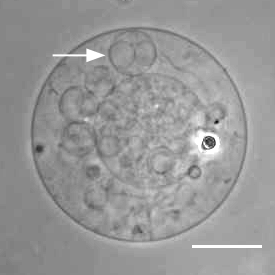}}
   \subfigure[]{\includegraphics[width=0.49\textwidth]{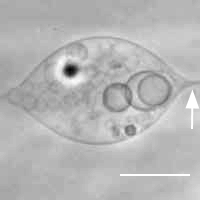}}  %% Zoomed out??
   \subfigure[]{\includegraphics[width=0.49\textwidth]{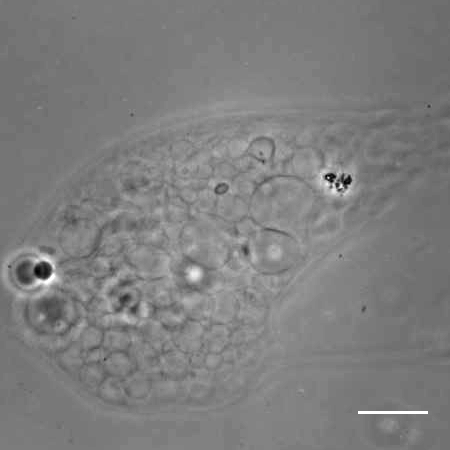}}
    \caption{Photomicrographs to show the four varieties of MVL identified. (a) Type 1a. (b) Type 1b. An internal liposome containing recursive liposomes is arrowed. (c) Type 2. Stalk is arrowed. (d) Type 3. All scale bars 50~$\mu$m.}
    \label{fig-types}
\end{figure}

The following basic varieties of MVLs were observed (Fig. \ref{fig-types}):
\begin{description}[labelindent=1cm]
    \item[$\rightarrow${\bf Type 1a}] Spherical liposomes, containing smaller liposomes. 72\% \linebreak prevalence.
    \item[$\rightarrow${\bf Type 1b}] As in 1a, but a further degree of recursion was observed, i.e. liposomes within liposomes within liposomes. 12\% prevalence.
    \item[$\rightarrow${\bf Type 2}] Spherical liposome with pinched ends, possessing a thin \linebreak stalk-like structure anchoring it to a dried lipid spot. 8\% prevalence.
    \item[$\rightarrow${\bf Type 3}] Amorphous structure, typically very large and possessing a stalk as per type 2. 8\% prevalence.
\end{description}

All varieties of MVL were observed in both swelling and budding phases of electroformation and also after electroformation had completed. The highest degree of recursion observed was 3, or more plainly, we observed MVLs that contained at least one inner compartment, which in turn contained at least one inner compartment. Type 1a, 1b and 2 MVL sizes ranged from 17--174~$\mu$m, mean 65~$\mu$m (see Table \ref{table1}) and type 3 vesicles ranged from 70--$246~\mu$m, mean 140~$\mu$m, when measured along their maximal dimensions. These two groups were conceptually split due to the amorphous nature of type 3 MVLs disallowing for direct comparison with other types.

Of the Type 1a, 1b and 2 MVLs observed, the number of internal recursive liposomes ranged between 1--14, mean 5 (see Table \ref{table1}), although several observations were omitted due to their contents being too unclear to record and/or containing membrane folds that did not constitute whole vesicles. Further, this number was not representative of internalised liposomes that were outside of the plane of view in the data collected. In all type 3 MVLs observed, the vesicle count was too numerous to accurately count. 

MVLs were found to be stable in solution for at least 5 days when stored in a sealed, refrigerated container (4$^o$C).

\subsection{Symbol embedding}
\begin{figure}
    \centering
    \includegraphics[width=0.5\textwidth]{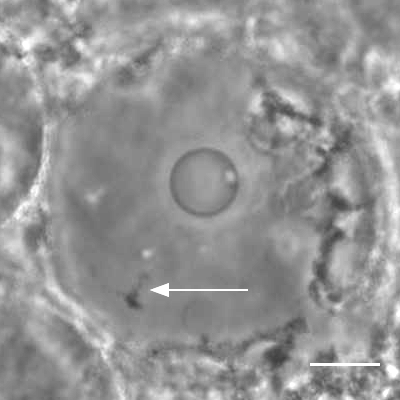}
    \caption{Photomicrograph to show a type 1a MVL containing a smaller liposome and a nanoparticle aggregate (arrowed). The periphery of the external liposome is coated with further nanoparticle aggregates. Scale bar 50~$\mu$m.}
    \label{fig-magnetiteLiposome}
\end{figure}

Objects with an appearance consistent with aggregated magnetite nanoparticles were observed within MVLs in experiments where they had been added to the swelling solution (Fig. \ref{fig-magnetiteLiposome}). In all experiments (n=15), nanoparticle aggregates were only observed in the outer-most compartment of MVLs, although the imaging technique did not resolve individual nanoparticles. This method typically resulted in all MVLs being coated in adherent nanoparticle aggregates and facilitated their being physically manipulated by a permanent magnet held in close proximity.

Continuous AC stimulation with low frequency signals ($\leq$2~Hz) caused observable oscillation in all liposomes that were adherent to the glass substrate. Only the external compartment membrane oscillated in phase with the applied signal. MVLs containing a small number of internal compartments were observed to cause sympathetic, out-of-phase mechanical oscillation in their internal compartment/s; this effect was not observed in MVLs that contained, tightly packed internal liposomes (See Supplementary Information, Video 1).

\begin{figure}
    \centering
    \subfigure[0~ms]{\includegraphics[width=.49\textwidth]{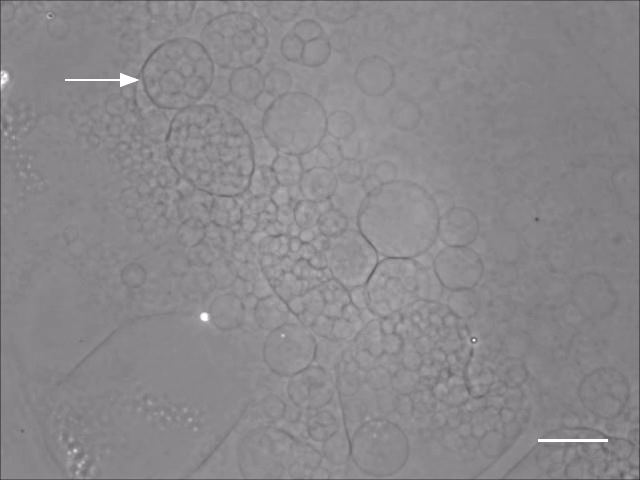}}
        \subfigure[1~ms]{\includegraphics[width=.49\textwidth]{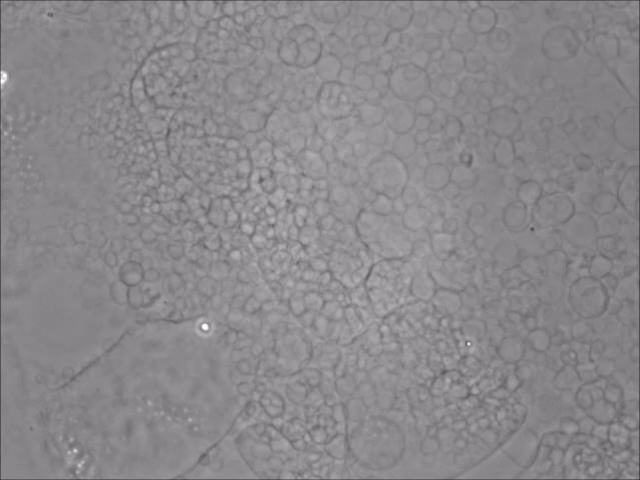}}
    \subfigure[2~ms]{\includegraphics[width=.49\textwidth]{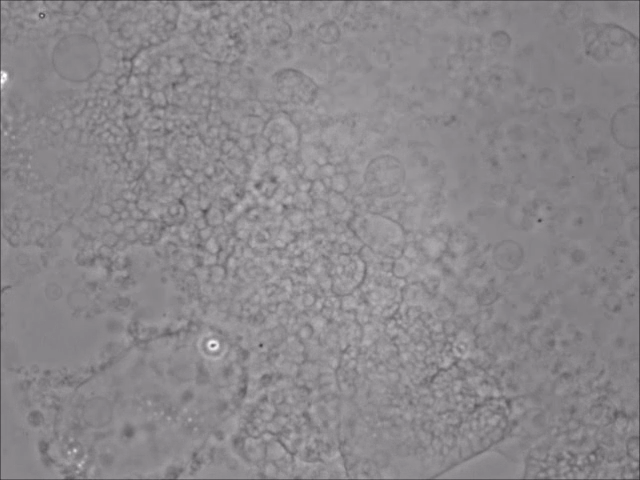}}
    \caption{Videomicrograph stills to show lysis of MVL external membranes. Several MVLs (one type 1a is arrowed in [a]) are shown at 1~ms intervals after the application of a brief DC pulse. The external membranes can be observed to burst and liberate their contents. Scale bar 50~$\mu$m.}
    \label{fig-lysis}
\end{figure}

Application of DC pulses caused rapid lysis of liposomes and MVLs. Application of brief DC pulses (20~V RMS as a 1~Hz pulse, pulse width 200~$\mu$s, rise 17~ns) was found to lyse MVL external compartments and liberate the internal ones undamaged (Fig. \ref{fig-lysis}).

\begin{table}
    \centering
    \begin{tabular}{c|c|c|c}
             & MVL Size, T1a, 1b, 2 & MVL Size, T3 & No. of Internal Liposomes\\
             \hline
             \hline
    Mean     & 64.60      & 127.81 & 5.10               \\
    St. Dev. & 40.19      & 70.96 & 3.15               \\
    Range    & 17.39--173.50    & 69.97--246.18 &    1--14           \\
    \end{tabular}
    \caption{Table to show collated data for physical characteristics of MVLs. MVL sizes are measured in $\mu$m; type 3 MVLs were measured across maximal dimension. Internal liposomes were manually counted. Type 1--2 size n=54, internal liposome count n=49, type 3 size n=5.}
    \label{table1}
\end{table}

\section{Discussion}
\subsection{Practical aspects and MVL characteristics} 
Here we described a modification of a well-known method for standard liposome production that may be used to robustly form MVLs. MVLs tended to form during early stages of electroformation, which is consistent with the theory that they form as intermediates in the generation of unilamellar liposomes: the methods presented here have the benefit of allowing for MVL formation to be observed and stopped at the most opportune stage. MVLs could also be kept adherent to their glass substrate by removing electrical stimulation (hence omitting the budding phase at a lower frequency) during the early stages of electroformation, therefore facilitating their micromanipulation.

Another key benefit of these methods is that the experiment is frugal, versatile and simple, hence it is well suited for its adoption as a method in unconventional computing research. Unconventional computing is the search for new materials, applications and uses for computing technologies that necessarily draws from a massively multidisciplinary base. 

When considered as an unconventional computer prototype, a typical MVL P-system is approximately 65~$\mu$m in diameter and contains about 0.1~mL of fluid. The number of internal compartments and their respective volumes is of course variable, but this is unimportant when considering that the reactions designed within are bounded by the level of recursion and the massive parallelism at which they can take place. To comment on the efficiency of putative MVL P-systems, each is produced in an experiment that draws approximately 12~$\mu$W of electrical power, benefits from possessing millions of `processors' (i.e. reactant molecules, the number of which is bounded by Avogadro's constant) and is highly unlikely to generate excessive waste energy thermalisation in the course of their operation at room temperature in an aqueous environment. 

The experimental setup used is reasonably easy to assemble in a properly-equipped laboratory and cheap to run per experiment: we estimate each electroformation chamber costs approximately 10GBP per experiment (see Supplementary Information Sheet 1). MVL formation should therefore be comparatively accessible to Computer Scientists who do not have a background in the life sciences.

The resilience studies performed here indicate that MVLs are stable for several days, which is long enough to permit extensive experimentation. We found the maximum level of recursion exhibited by MVLS produced through these methods to be 3-layers. Size of MVL did not seem to correlate with the number or size of internal liposomes. %STATS?

%% Separation methods? > Harvesting

\subsection{Embedding symbols and reactions}

Embedding chemical reactions within liposomes has been a topic of study for several decades and great advances have been made, such as self-assembly of nanoscale protein networks inside liposomes \cite{Pontani2009} and embedding of functional transmembrane transporter proteins in their membranes \cite{Paternostre1988,Simeonov2013}. 

In this subsection, we collate both well known mechanisms for manipulating synthetic liposomes and their contents as well as the novel principles of introducing symbols into multivesicular systems described in this study, for the purpose of discussing how specific experimental P-systems may be implemented.

{\bf 1: Diffusion}\\
Phospholipid membranes are permeable to small, uncharged polar molecules (e.g. water, urea) and lipophilic molecules (steroids, phospholipid analogues).

Whilst introducing diffusable chemical reactants to MVL systems post-formation is simple, this limits the direction of the reaction from outer compartments to inner.

{\bf 2: Introduction via swelling solution}\\
Addition of chemicals into the swelling solution is a method towards embedding symbols in equal concentration between recursive compartments, e.g. we assume that all MVL compartments generated in the experiments above will contain sucrose at a concentration of 300mM, which could be used to represent symbols. We also experimented with embedding micro-scale aggregated magnetite nanoparticles into growing MVLs via their inclusion in the swelling solution, but this was found to result in their distribution in MVL outer components only (Fig. \ref{fig-magnetiteLiposome}). Hence, differential distribution of reactants between layers may be achieved through using labelled nano- and microparticles.

Again, this method would be a simple route towards symbol embedding, although the range of chemicals that may be used for this purpose are limited by their effects on phospholipid membranes. In initial experiments to implement chromogenic reactions in MVLs (data not shown), addition of cobalt chloride was found to prevent liposome formation. Further, high ionic strength solutions prevent proper hydration of the lipid layer \cite{Estes2005}.

{\bf 3: Influencing membrane permeability}\\
Membrane permeability may be manually altered through the use of microinjection, electroporation and embedding of transmembrane channel proteins. All of these techniques are technically complex and require specialised equipment, but are well established in biology laboratories for inserting substances into membrane-encapsulated systems. 

Microinjection of substances into membrane-encapsulated compartments via a hollow glass needle  allows for virtually any lipid membrane-compatible compound to be injected into a MVL compartment. Electroporation involves using a rapid, high voltage pulse to perforate PL membranes and is frequently used to introduce DNA fragments into individual cells, hence this represents a route towards altering the outer component membrane permeability whilst leaving internal membranes intact. Insertion of transmembrane proteins into an experimental P-system would allow a MVL to selectively permit the bi-directional flow of ions in a manner similar to live cells and therefore represents a highly desirable, if technically complex, objective.

{\bf 4: Unconventional symbols}\\
Data in P-systems are typically considered to constitute chemical reactants that possess their own rule set. If data are considered in a more abstract sense, however, there exist further possibilities for representing data within MVLs.

For example, embedding of ferromagnetic nanoparticles would offer the prospect of using magnetic localisation of compartments or magnetic data storage via a read/write head. Both of these examples could be simultaneously coupled with conventional chemical reactions, e.g. reaction of ferric iron with potassium ferrocyanide to complete a chromogenic reaction to produce Prussian blue. 

Another possibility is the use of electromagnetic radiation as an input for inducing vibrations in membranes of MVLs adherent to their substrate and for spontaneously destroying membranes (i.e. erase/halt operations) when applied as DC. Comparing the induced frequency of motion between external and internal compartments of a MVL when AC signals are applied could be considered to function in a manner similar to an analogue computer.

{\bf Example of a possible P-system implementation}\\
Consider the preparation of MVLs in an aqueous sucrose solution containing the enzyme urease, which is embedded into all of its compartments. After formation, the MVLs are removed and transferred to fresh media. Urea is then added to the media, which passively diffuses through the MVLs' external membrane and hydrolyses in the presence of its enzyme catalyst, as per the equation below. The reaction liberates carbon dioxide gas and ammonia. Fine-tuning of urease and urea concentrations (i.e. input data) will dictate whether sufficient unhydrolysed urea passes through the MVLs' outer-most compartments into its inner compartments. Computation can be continuously monitored by optically assessing the comparative sizes of compartments due to the effects of swelling from evolved carbon dioxide, or through observing colourimetric changes with a pH indicator dye. The process would necessarily be halting due to end conditions of reactant depletion and/or bursting of compartments due to the evolution of carbon dioxide.

\begin{equation}
    (NH_2)_2CO + H_2O \rightarrow CO_2 + 2NH_3
    \label{eq1}
\end{equation}

\section{Conclusion}
We have described methods by which experimental P-systems may be created in a laboratory setting and produced a basic characterisation of their physical characteristics, including size and apparent boundaries for number of internal compartments. Although further work is required before it will be apparent how `useful' experimental P-systems could be, there are a range of potential uses for multi-compartment massively-parallel bioreactors, as indicated by previous theoretical work in the field. Generation of `real' P-systems presents significant technical challenges in their construction and programming, however: we recommend that further work focus on making protocols for robust, fully automated MVL formation and evaluation of the methods towards symbol embedding described in the previous section.

\section*{Declaration of Interest}
The authors declare no conflict of interest.

\section*{Appendix: Supplementary Information}
\noindent
Sheet 1: Materials list.\\
Video 1: Video to show several MVLs physically oscillating under the influence of a 2~Hz, 1.2~Vpp sine wave. All liposomes are adherent to the base of an ITO-coated glass slide as the video footage was collected at the start of the `budding' phase. Several type 1a MVLs are present, one of which is arrowed.

\bibliographystyle{spphys}
\bibliography{liposomesReferences}

\end{document}